\documentclass[journal=jpccck,manuscript=article,layout=twocolumn]{achemso}

\usepackage[version=3]{mhchem} 

\author{Seho Yi}
\affiliation[HYU]
{Department of Physics and Research Institute for Natural Sciences, Hanyang University, 222 Wangsimni-ro, Seongdong-gu, Seoul 133-791, Korea}
\author{Hyun-Jung Kim}
\affiliation[KIAS]
{Korea Institute for Advanced Study, 85 Hoegiro, Dongdaemun-gu, Seoul 130-722, Korea}
\author{Jin-Ho Choi}
\email{jinogino@hanmail.net}
\affiliation[RIMT]
{Research Institute of Mechanical Technology, Pusan National University, Busandaehang-ro 63 beon-gil 2, Geumjeong-gu, Busan 609-735, Korea}
\author{Jun-Hyung Cho}
\email{chojh@hanyang.ac.kr}
\affiliation[HYU]
{Department of Physics and Research Institute for Natural Sciences, Hanyang University, 222 Wangsimni-ro, Seongdong-gu, Seoul 133-791, Korea}

\title[\textsf{achemso}]{Contrasting diffusion behaviors of O and F atoms on graphene and within bilayer graphene}

\begin{document}


\begin{tocentry}

\centering
  \includegraphics{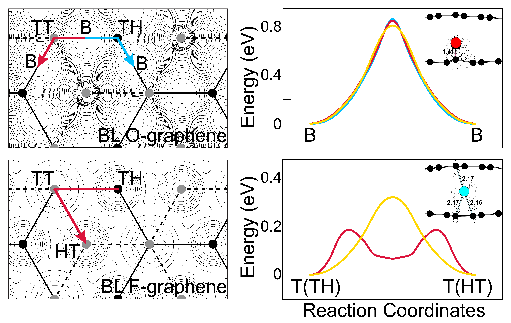}
\end{tocentry}

\begin{abstract}
The chemical modification of graphene with adatoms is of importance for nanoelectronics applications. Based on first-principles density-functional theory calculations with including van der Waals interactions, we present a comparative study of the diffusion characteristics of oxygen (O) and fluorine (F) atoms both on graphene and between the layers of bilayer graphene. We find that O and F atoms have lower binding energies between the layers of bilayer graphene compared to on graphene. Interestingly, the calculated diffusion barrier for the O atom slightly increases from 0.81 eV on graphene to 0.85 eV within bilayer graphene, while that for the F atom significantly decreases from 0.30 eV on graphene to 0.18 eV within bilayer graphene. Such contrasting behaviors of the O and F diffusions within bilayer graphene can be traced to their different bonding natures: i.e., the O adatom that has a strongly covalent C$-$O$-$C bonding on the bridge site of the C$-$C bond diffuses on one graphene layer with a slight interference of the other layer, while the F adatom that has a weakly ionic F$-$C bonding on top of a C atom easily diffuses by hopping between two graphene layers with accepting more electron charges from the two layers. The present findings have important implications for understanding the diffusion processes of F and O atoms on graphene and within bilayer graphene.
\end{abstract}


\vspace{0.5cm}

\section{Introduction}
Since the discovery of graphene in 2004,~\cite{Nov} its application to nanoelectronic devices has been regarded as one of the most significant tasks in the field of nanotechnology sciences. However, the absence of a band gap in graphene restricts its utilization as a channel material in field-effect transistors. So far, many efforts have been devoted to open a band gap in graphene.~\cite{Gio,Han,Gui,Zan,Lu,Sof,Gom,Jung,Yan,Char,Chen,Med,Kar} Specifically, the chemical modification approach with adatoms~\cite{Sof,Gom,Jung,Yan,Char,Chen,Med,Kar} has been intensively employed for the gap opening of graphene. For instance, fluorinated graphene (hereafter termed F-graphene) has been attracted much attention as a promising material for nanoelectronic device applications because of its tunable band gap with respect to the amount of fluorination and the fluorination patterns.~\cite{Geo,Rob,Wang} In order to design the structure of F-graphene in atomic scale, it is necessary to understand the diffusion and adsorption of the F atom on graphene. It was found that the F atom prefers to adsorb on the on-top site of graphene with a single F$-$C bond.~\cite{Weh,Are} Recently, Sadeghi $et$ $al$.~\cite{Sad} reported a first-principles density-functional theory (DFT) study of the F diffusion both on monolayer (ML) graphene and between the layers of bilayer (BL) graphene, showing a drastic difference between the two cases: i.e., the mobility of the F atom within BL graphene increases by about an order of magnitude compared to that on ML graphene. Although Sadeghi $et$ $al$.~\cite{Sad} predicted a facilitated diffusion process of the F atom within BL graphene by hopping from one layer to the other layer, the microscopic underlying mechanism of the drastic difference of F diffusion between ML and BL graphenes is still lacking.

As another chemical modification of graphene, oxidized graphene (hereafter termed O-graphene) has been widely adopted because of its low cost and manufacturing feasibility for mass production of graphene or graphene nano-platelets.~\cite{Dre} In particular, O-graphene which can be generated from graphite oxide contains various O-containing functional groups, thereby offering a broad applicability to electronic devices. It is noted that the O atom adsorbs on the bridge site of graphene with the C$-$O$-$C bond,~\cite{Yan,Weh,Sua,Top,Yaf} which is different from the on-top adsorption site of hydrogen, halogen, and F atoms.~\cite{Sof,Char,Med,Sad,Weh} This different binding of O with graphene may cause a distinctive diffusion behavior compared to the above-mentioned F diffusion. Indeed, earlier DFT calculations~\cite{Weh,Sua,Top,Yaf} reported that the energy barrier for the O diffusion on ML graphene ranges between 0.73 and 0.81 eV, which is relatively larger than that (ranging between 0.13 and 0.29 eV) for the F diffusion on ML graphene.~\cite{Sad,Weh,Are} However, a comparative study of the O diffusions on ML graphene and within BL graphene is yet to be explored.

\begin{figure*}[ht]
\includegraphics[width=16cm]{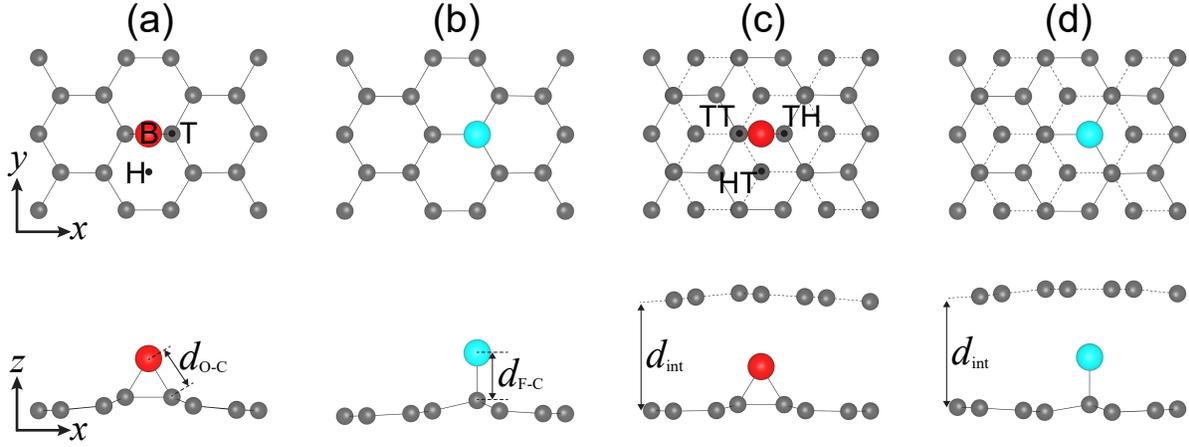}
\caption{Top and side views of the structures of the (a) O and (b) F atoms on ML graphene, optimized using the PBE+vdW calculation. The corresponding ones of the O and F atoms within BL graphene are drawn in (c) and (d), respectively. For distinction, the C$-$C bonds in the upper layer of BL graphene are drawn with the dashed lines. On ML graphene, B, T, and H denote the bridge, top, and hollow sites, respectively. Meanwhile, within BL graphene, TH, TT, and HT denote the top(in lower graphene sheet)-hollow(in upper graphene sheet), top-top, and hollow-top sites, respectively.}
\label{figure:1}
\end{figure*}

In this paper, we perform a first-principles theoretical study of the diffusion behaviors of the O and F adatoms both on ML graphene and between the layers of BL graphene by using a van der Waals (vdW) energy-corrected DFT calculation.~\cite{TS} Our calculated potential energy surfaces (PESs) for such ML and BL O(F)-graphene systems show that (i) the diffusion barrier $D_b$ for O (F) on ML graphene is 0.81 (0.30) eV, and (ii) $D_b$ for O (F) within BL graphene increases (decreases) to 0.85 (0.18) eV. Such contrasting variations of the O and F diffusions between ML and BL graphenes can be associated with their different bonding natures: i.e., the O adatom has a strongly covalent C$-$O$-$C bonding on the bridge site of the C$-$C bond, while the F atom has a weakly ionic F$-$C bonding on top of a C atom. Consequently, the O atom diffuses on one graphene layer with a slight interference of the other layer, while the F atom easily diffuses by hopping between two graphene layers with accepting more electron charges from the two layers. Our findings shed more light on understanding the contrasting diffusion behaviors of the O and F atoms on ML graphene and within BL graphene.

\section{Computational Details}
Our first-principles DFT calculations were performed using an accurate all-electron, full-potential scheme with numeric atom-centered orbital basis functions, as implemented in the Fritz-Haber-Institute $ab$ $initio$ molecular simulations (FHI-aims) package.~\cite{aims} All the calculations were carried out with ``tight'' computational settings and accurate tier-2 basis sets. The generalized gradient approximation (GGA) functional of Perdew-Burke-Ernzerhof (PBE)~\cite{PBE} was employed for the exchange-correlation energy. To include the effects of vdW interactions, we used the PBE+vdW scheme developed by Tkatchenko and Scheffler,~\cite{TS} which has been demonstrated to be accurate in layered crystals as well as various adsorbates on surfaces.~\cite{Ruiz,Gan} The O-graphene or F-graphene system was simulated by containing one O or F atom within a $3\sqrt{3}$${\times}$$3\sqrt{3}$ unit cell, thereby minimizing the spurious interactions between O or F adatoms in the periodic unit cells. The \textbf{k}-space integration was done with the 12${\times}$12 uniform meshes in the two-dimensional Brillouin zone. All atoms were allowed to fully relax until all the residual force components on each atom were less than 0.01 eV/{\AA}. Using the PBE (PBE+vdW) calculation, the lattice constant of ML graphene and the interlayer distance of BL graphene were obtained as 2.47 (2.46) and 4.06 (3.24) {\AA}, respectively.

\begin{table}[ht]
\caption{Calculated PBE binding energies ($E_b$) of the O-graphene and F-graphene systems, together with the PBE+vdW results in parentheses\textsuperscript{\emph{a}}}

{\renewcommand{\arraystretch}{1.2}
\begin{tabular}{ccclc}
&              &\multicolumn{2}{c}{$E_{\rm b}$ (eV/atom)} \\
              \hline
&ML O-graphene &B & 2.35 (2.41) &\\
&              &T & 1.53 (1.60) &\\
&              &H & 0.57 (0.64) &\\
&              &B & 2.43\textsuperscript{\emph{b}} &\\
&ML F-graphene &T & 1.88 (1.95) &\\
&              &H & 1.45 (1.48) &\\
&              &T & 1.99\textsuperscript{\emph{b}} &\\
&BL O-graphene &B & 2.24 (1.75) &\\
&              &TT& 1.50 (0.90) &\\
&              &TH& 1.47 (0.89) &\\
&BL F-graphene &TH& 1.81 (0.97) &\\
&              &TT& 1.67 (0.90) &\\
&              &TH& (1.18)\textsuperscript{\emph{c}} &\\
              \hline
\end{tabular}
}
\\
\textsuperscript{\emph{a}} $E_b$ is defined as $E_{\rm tot}$(graphene) + $E_{\rm tot}$(X) $-$ $E_{\rm tot}$(X-graphene), where $E_{\rm tot}$(X-graphene), $E_{\rm tot}$(graphene), and $E_{\rm tot}$(X) denote the total energies of O-graphene or F-graphene, pristine ML or BL graphene, and free O or F atom, respectively. \\
\textsuperscript{\emph{b}} Ref. 19: PBE. \\
\textsuperscript{\emph{c}} Ref. 20: PBE with vdW interactions.

\end{table}

\section{Result and Discussion}

We first perform the PBE calculation to find the optimized structures and binding energies of the O and F adatoms on ML graphene and between the two layers of BL graphene. The PBE results for the binding energies ($E_b$) of all the considered adsorption sites (see Figure 1) are summarized in Table 1. We find that $E_b$ is 2.35, 1.53, and 0.57 eV for the B, T, and H sites of ML O-graphene; 2.24, 1.50, and 1.47 eV for the B, TT, and TH sites of BL O-graphene, respectively. Meanwhile, $E_b$ is calculated to be 1.88 and 1.45 eV for the T and H sites of ML F-graphene; 1.81 and 1.67 eV for the TH and TT sites of BL F-graphene, respectively. Therefore, for ML O-graphene, the O (F) atom prefers to adsorb at the B (T) site, consistent with previous DFT calculations.~\cite{Weh,Sad,Sua} It is noted that the difference in the adsorption sites of the O and F atoms can be attributed to the so-called octet rule: i.e., the O atom has a strongly covalent bonding with two C atoms on the bridge site of the C$-$C bond, while the F atom has a weakly ionic bonding to a single C atom on top of a C atom. These different bonding natures of the O and F adatoms are demonstrated by their charge character analysis, as discussed below.

\begin{table}[ht]
\caption{Calculated PBE bond lengths ($d_{\rm X-C}$: X = O, F) and interlayer distances ($d_{\rm int}$) of the O-graphene and F-graphene systems, together with the PBE+vdW results in parentheses\textsuperscript{\emph{a}}}
{\renewcommand{\arraystretch}{1.6}
\begin{tabular}{ccc}
              &$d_{\rm X-C}$ ({\AA})&$d_{\rm int}$ ({\AA}) \\ \hline
ML O-graphene & 1.46 (1.46) & $-$ \\
ML F-graphene & 1.54 (1.54) & $-$ \\
BL O-graphene & 1.46 (1.44) & 4.23 (3.49) \\
BL F-graphene & 1.59 (1.53) & 4.75 (3.55) \\
\hline
\end{tabular}
}
\\
\textsuperscript{\emph{a}} For the definitions of $d_{\rm X-C}$ and $d_{\rm int}$, see Figure 1.
\end{table}

Due to the interlayer vdW interactions in BL graphene, the geometries and binding energies of the BL O-graphene and BL F-graphene systems are expected to be different between the PBE and PBE+vdW results. To examine these differecnes, we perform the PBE+vdW calculation for various adsorption sites. Figure 1a-d show the most stable structures of ML O-graphene, ML F-graphene, BL O-graphene, and BL F-graphene, respectively, obtained using the PBE+vdW calculation. The calculated binding energies of these O-graphene and F-graphene systems are also listed in Table 1. We find that the most stable B site for ML or BL O-graphene and T (TH) site for ML (BL) F-garphene are invariant between the PBE and PBE+vdW calculations. However, the inclusion of vdW interactions in ML O-graphene and ML F-graphene slightly increases $E_b$ by 0.06 and 0.07 eV, respectively. On the other hand, such vdW effects in BL O-graphene and BL F-graphene decrease $E_b$ as large as 0.49 and 0.84 eV, respectively (see Table 1). This huge vdW-induced reduction of binding energy in BL O-graphene and BL F-graphene is caused by the fact that the inclusion of vdW interactions produces the large lattice deformation of BL graphene. As shown in Table 2, for ML O-graphene and ML F-graphene, the PBE and PBE+vdW calculations hardly change the bond lengths $d_{\rm O-C}$ and $d_{\rm F-C}$. However, the PBE+vdW geometry of BL O(F)-graphene shows a dramatic decrease in the interlayer distance $d_{\rm int}$ by 0.74 (1.20) {\AA} compared to the corresponding values obtained using PBE (see Table 2). Consequently, the PBE+vdW calculations for BL O-graphene and BL F-graphene give rise to a significant structural deformation of the upper graphene layer around O and F adatoms [see Figure 1c,d], thereby leading to a decrease in their binding energies compared to those computed from PBE. In these respects, we can say that vdW interactions play an important role in determining the geometries and binding energies of BL O-graphene and BL F-graphene.

Next, we investigate the diffusions of O and F atoms on ML graphene using the PBE+vdW scheme. To obtain the minimum-energy diffusion pathway, we calculate the PESs for ML O-graphene and ML F-graphene by optimizing the structure at the uniformly separated adsorption sites within the graphene unit cell. Figure 2a,b) show the calculated PESs for ML O-graphene and ML F-graphene, respectively. We find that, for ML O-graphene, the O adatom has a diffusion path from the B site to the neighboring B site through the T$'$ (close to T) site with $D_b$ = 0.81 eV, in good agreement with a previous DFT result~\cite{Sua} where the transition state is slightly deviated from the T site with $D_b$ = 0.73 eV. Here, the present angles of the C$-$C$-$O bonds at the T$'$ site are 101.5$^{\circ}$, 102.5$^{\circ}$, and 106.3$^{\circ}$ (see Figure S1 of the Supporting Information), close to those (99.9$^{\circ}$, 101.0$^{\circ}$, and 108.1$^{\circ}$) of a previous DFT calculation.~\cite{Sua} Meanwhile, for ML F-graphene, the F adatom diffuses from the T site to the neighboring T site through the B site with $D_b$ = 0.30 eV, in good agreement with that (0.29 eV) of previous DFT calculations.~\cite{Weh,Are} Thus, $D_b$ for the O diffusion of ML O-graphene is $\sim$2.5 times larger than that for the F diffusion of ML F-graphene, indicating that the F atom can diffuse much faster than the O atom.

\begin{figure}[h!t]
\includegraphics[width=8cm]{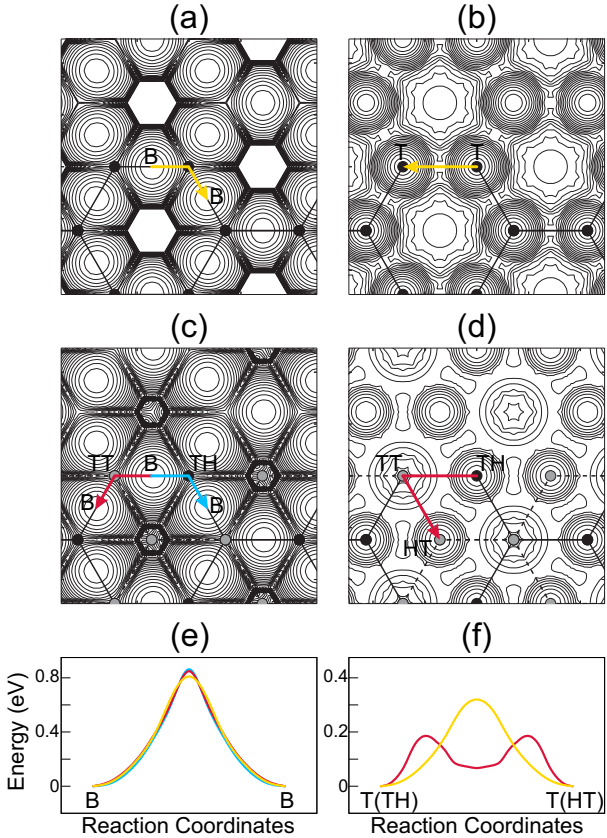}
\caption{Calculated PESs for (a) ML O-graphene, (b) ML F-graphene, (c) BL O-graphene, and (d) BL F-graphene. The contour spacing is 0.1 and 0.025 eV for the O-graphene and F-graphene systems, respectively. The C-C bonds in the lower and upper layers of BL graphene are drawn with the solid and dashed lines, respectively. The energy profiles along the diffusion paths of O-graphene and F-graphene, indicated by the arrows in (a), (b), (c), and (d), are given in (e) and (f), respectively. The energy zero is set to the equilibrium state for each path. }
\label{figure:2}
\end{figure}

To understand the contrasting diffusion behaviors of the O and F atoms between ML O-graphene and ML F-graphene, we examine the bonding natures of the O and F adatoms on ML graphene. Figure 3a,b display the charge density difference calculated from ML O-graphene and ML F-graphene, respectively, which is defined as
\begin{equation}
{\Delta}\rho = {\rho}_{\rm X-graphene} - ({\rho}_{\rm X} + {\rho}_{\rm graphene}).
\end{equation}
Here, ${\rho}_{\rm X-graphene}$ is the charge density of O-graphene or F-graphene and ${\rho}_{\rm X}$+${\rho}_{\rm graphene}$ is the superposition of the charge densities of the separated systems, i.e., isolated O or F atom and clean graphene. In Figure 3a, ${\Delta}{\rho}$ of ML O-graphene represents a covalent bonding character for the C$-$O$-$C bond with $d_{\rm O-C}$ = 1.46 {\AA}. Along the diffusion path B${\rightarrow}$T$'$${\rightarrow}$B, one O$-$C bond is broken while the other O$-$C bond is shortened to be $d_{\rm O-C}$ = 1.41 {\AA} at the transition site. Meanwhile, as shown in Figure 3b, ${\Delta}{\rho}$ of ML F-graphene represents an ionic bonding character for the F$-$C bond with $d_{\rm F-C}$ = 1.54 {\AA}. Note that F has a relatively larger electronegativity of 3.98 compared to that (2.55) of C.~\cite{all} This ionic F$-$C bond character for an isolated F adatom on graphene is consistent with a recent combined DFT and X-ray photoelectron spectroscopy study.~\cite{ionic} Interestingly, the F atom at the transition state (equivalently at the B site) along the diffusion path T${\rightarrow}$B${\rightarrow}$T is far away from two neighboring C atoms with $d_{\rm F-C}$ = 2.29 {\AA}. This longer separation between the F and C atoms at the transition state is drastically different from a decrease of $d_{\rm O-C}$ at the transition state of ML O-graphene, reflecting the different bonding natures between ML O-graphene and ML F-graphene. As shown in the right panels of Figure 3a,b, ${\Delta}{\rho}$ for the transition states of ML O-graphene and ML F-graphene clearly show the covalent and ionic bonding characters, respectively.

We continue to investigate the diffusion of the O and F atoms between the layers of BL graphene using the PBE+vdW scheme. Figure 2c,d display the calculated PESs for BL O-graphene and BL F-graphene, respectively. We find that the O atom in BL O-graphene is covalently bound to one graphene layer during the diffusion process. Compared with the result for ML O-graphene, the O diffusion in BL O-graphene is somewhat influenced by the presence of the other layer. For BL O-graphene, there are the two different diffusion paths I and II with $D_b$ = 0.85 and 0.86 eV, respectively, which is slightly larger than (0.81 eV) in ML O-graphene. Here, the path I (II) has the transition state passing the TT (TH) site [see Figure 2c]. Note that the hopping of the O atom from one layer to the other layer has a high energy barrier of ${\sim}$1.3 eV. By contrast, the PES for BL F-graphene shows that the F atom easily diffuses by hopping from one layer to the other layer: see Figure 2d. We find that the diffusion path along TH${\rightarrow}$B${\rightarrow}$TT${\rightarrow}$B${\rightarrow}$HT involves double barriers with $D_b$ = 0.18 eV  [see Figure 2f]. This value in BL F-graphene is much reduced compared to that (0.30 eV) in ML F-graphene, indicating that the diffusion of F is facilitated within BL graphene. These contrasting diffusion behaviors of O and F between BL O-graphene and BL F-graphene can be also traced to their different bonding natures, as discussed below.

\begin{figure}[h!t]
\includegraphics[width=8.cm]{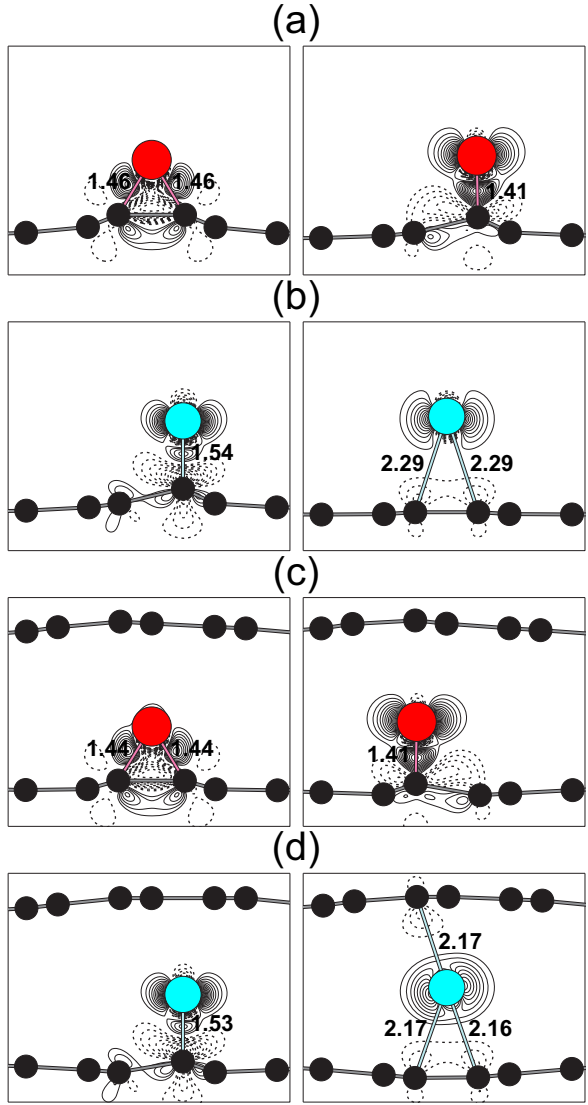}
\caption{${\Delta}{\rho}$ of (a) ML O-graphene, (b) ML F-graphene, (c) BL O-graphene, and (d) BL F-graphene. The left (right) panels represent the results at the equilibrium (transition) state. In the contour plots, the first solid (dashed) lines are drawn at 5.0${\times}$10$^{-3}$ $e$/{\AA}$^3$ ($-$5.0${\times}$10$^{-3}$ $e$/{\AA}$^3$) and the contour spacing is 3.3${\times}$10$^{-3}$ $e$/{\AA}$^3$ ($-$3.3${\times}$10$^{-3}$ $e$/{\AA}$^3$). The solid and dashed lines indicate accumulated and depleted electrons, respectively. The numbers represent the bond lengths (in {\AA}).}
\label{figure:3}
\end{figure}

As shown in Figure 3a,c, the values of $d_{\rm O-C}$ at the equilibrium and transition states of BL O-graphene are similar to those in ML O-graphene, implying that O adsorption on one layer is little affected by the other layer. Such small changes of the geometries of ML O-graphene and BL O-graphene is consistent with their similarity of ${\Delta}{\rho}$: see Figure 3a,c. These results in turn give the slight change in $D_b$ between ML O-graphene and BL O-graphene. On the other hand, $d_{\rm F-C}$ at the transition state of BL F-graphene becomes 2.16 or 2.17 {\AA}, much shorter than that (2.29 {\AA}) in ML F-graphene: see the right panels of Figure 3b,d. This decrease of $d_{\rm F-C}$ in BL F-graphene can be associated with more ionic character due to its enhanced charge transfer: i.e., the F adatom in BL F-graphene accepts more electrons simultaneously from the two layers compared to ML F-graphene, as shown in Figure 3b,d. Such more charge accumulation around the F atom is likely to increase the binding energy due to an ionic bonding to its neighboring C atoms, leading to a decrease of $D_b$ in BL F-graphene.

It is finally worth noting that the adsorption of alkali and transition metal elements on graphene occurs at the H site, forming strong chemical bonds with surrounding C atoms.~\cite{Man,Yao} Such strong bindings on graphene cannot be expected to an easy diffusion. Indeed, some transition metal elements favoring adsorption on the H site showed relatively larger diffusion barriers compared to other elements favoring adsorption on the B or T site~\cite{Man}. It is thus likely that the initial adsorption structure of adatoms on graphene plays a crucial role in determining their diffusion behaviors.

\section{Conclusion}
We have presented a first-principles theoretical study of the diffusion behaviors of the O and F adatoms both on ML graphene and between the layers of BL graphene. Our calculated potential energy surfaces for the O-graphene and F-graphene systems showed that the diffusion barrier for the O atom slightly increases from 0.81 eV on graphene to 0.85 eV within bilayer graphene, while that for the F atom significantly decreases from 0.30 eV on graphene to 0.18 eV within bilayer graphene. It is revealed that such contrasting variations of the O and F diffusions between ML and BL graphenes are attributed to their different bonding natures: i.e., the O adatom that has a strongly covalent C$-$O$-$C bonding on the bridge site of the C$-$C bond diffuses on one graphene layer with a slight interference of the other layer, while the F adatom that has a weakly ionic F$-$C bonding on top of a C atom easily diffuses by hopping between two graphene layers with accepting more electron charges from the two layers. Our findings shed more light on understanding the contrasting diffusion behaviors of the O and F atoms on ML graphene and within BL graphene, which will provide a useful information for the chemical modification of graphene sheets.

\section{Associated Content}
\textbf{Supporting information}\\
The geometry of the transition state of ML O-graphene is given. This material is available free of charge via the Internet at http://pubs.acs.org.

\section{Author Information}
\textbf{Corresponding Author}\\
*Email: chojh@hanyang.ac.kr\\
*Email: jinogino@hanmail.net\\

\textbf{Notes}\\
The authors declare no competing financial interest.

\begin{acknowledgement}
This work was supported by National Research Foundation of Korea (NRF) grant funded by the Korean Government (2015R1A2A2A01003248) and KISTI supercomputing center through the strategic support program for the supercomputing application research (KSC-2016-C3-0001).
\end{acknowledgement}

\end{document}